\pdfoutput=1

\documentclass[3p]{elsarticle}

\usepackage{graphicx}
\usepackage{amssymb}
\usepackage{tabularx}
\usepackage{booktabs} 
\usepackage{siunitx}
\usepackage{placeins}
\usepackage{color}
\usepackage{soul}
\usepackage{textcomp}
\usepackage{hyperref}

\hypersetup{
	colorlinks   = true, 
	urlcolor     = blue, 
	linkcolor    = blue, 
	citecolor   = red 
}

\biboptions{sort&compress}

\journal{Polymer Degradation and Stability}

\begin{document}

\begin{frontmatter}

\title{Tuning structural relaxations, mechanical properties, and degradation timescale of PLLA during hydrolytic degradation by blending with PLCL-PEG}

\author[label1]{Reece N. Oosterbeek \corref{cor1}}
\ead{rno23@cam.ac.uk}

\author[label1]{Kyung-Ah Kwon}
\author[label2]{Patrick Duffy}
\author[label2]{Sean McMahon}
\author[label3]{Xiang C. Zhang}
\author[label1]{Serena M. Best}

\author[label1]{Ruth E. Cameron  \corref{cor1}}
\ead{rec11@cam.ac.uk}

\cortext[cor1]{Corresponding author}

\address[label1]{Cambridge Centre for Medical Materials, Department of Materials Science and Metallurgy, University of Cambridge, Cambridge, United Kingdom}
\address[label2]{Ashland Specialties Ireland Ltd, Synergy Centre, TUD Tallaght, Dublin, Ireland}
\address[label3]{Lucideon Ltd, Queens Road, Penkhull, Stoke-on-Trent, United Kingdom}

\begin{abstract}
Poly-\textsc{L}-lactide (PLLA) is a popular choice for medical devices due to its bioresorbability and superior mechanical properties compared with other polymers. However, although PLLA has been investigated for use in bioresorbable cardiovascular stents, it presents application-specific limitations which hamper device therapies. These include low toughness and strength compared with metals used for this purpose, and slow degradation. Blending PLLA with novel polyethylene glycol functionalised poly(\textsc{L}-lactide-co-$\varepsilon$-caprolactone) (PLCL-PEG) materials has been investigated here to tailor the mechanical properties and degradation behaviour of PLLA. This exciting approach provides a foundation for a next generation of bioresorbable materials whose properties can be rapidly tuned.

The degradation of PLLA was significantly accelerated by addition of PLCL-PEG. After 30 days of degradation, several structural changes were observed in the polymer blends, which were dependent on the level of PLCL-PEG addition. Blends with low PLCL-PEG content displayed enthalpy relaxation, resulting in embrittlement, while blends with high PLCL-PEG content displayed crystallisation, due to enhanced chain mobility brought on by chain scission, also causing embrittlement. Moderate PLCL-PEG additions (10\% PLCL(70:30)-PEG and 20 - 30\% PLCL(80:20)-PEG) stabilised the structure, reducing the extent of enthalpy relaxation and crystallisation and thus retaining ductility. Compositional optimisation identified a sweet spot for this blend strategy, whereby the ductility was enhanced while maintaining strength.

Our results indicate that blending PLLA with PLCL-PEG provides an effective method of tuning the degradation timescale and mechanical properties of PLLA, and provides important new insight into the mechanisms of structural relaxations that occur during degradation, and strategies for regulating these.

\end{abstract}


\end{frontmatter}


\section{Introduction}

As bioresorbable medical devices grow in popularity, poly-\textsc{L}-lactide (PLLA) has become a favoured material choice, especially for load bearing applications. This can be attributed to its ability to be completely resorbed into the body, as well as its high strength and stiffness compared with other biodegradable synthetic polymers. In particular it has gained prevalence in the field of bioresorbable cardiac stents, being used as the main stent structural material for commercial devices including the ABSORB (Abbott Vascular) and DESolve (Elixir Medical) stents \cite{McMahon2018}. These devices have shown promising initial results, demonstrating noninferiority to metallic stents, and low rates of complications \cite{Gao2015, Serruys2016a}, however more recent results have indicated increased rates of device-associated adverse events compared with metallic stents \cite{Kereiakes2017, Ali2018}. The ABSORB stent was voluntarily withdrawn from sale by the manufacturer in 2017 after the declaration of safety concerns, and to date the successful use of bioresorbable stents has been limited, in large part, by their mechanical and degradation properties \cite{McMahon2018, Mukherjee2017, Collet2016}.\\

Although PLLA is considered a relatively strong polymer, with yield strength of approximately 60 MPa and stiffness around 4 GPa \cite{Farah2016, NatureWorks2015a}, its mechanical properties are still insufficient for many load bearing applications, especially when compared with the permanent metals these polymers are often seeking to replace. For stents this means that to provide the required mechanical support, polymeric stents must incorporate significantly thicker struts than their metallic counterparts. Strut sizes below 100 $\mu$m, a common feature in metallic stents, have yet to be achieved by bioresorbable polymeric stents in a clinical setting. Thinner struts for bioresorbable polymeric stents are mandated based on the clinical evidence to date as they minimise blood flow turbulence, promote early endothelialisation, and reduce risks of thrombogenicity and restenosis \cite{McMahon2018, Kolandaivelu2011, Kastrati2001, Rittersma2004}. The slow degradation of PLLA is another issue that has been identified by clinical experts as requiring materials development - as a wound healing event, arterial patency is typically recovered within the first six months \cite{Rao2011, Nedeltchev2009}, however typical PLLA resorption can take several years \cite{Mukherjee2017, Serruys1988, Maurus2004, McMahon2018, Farah2016}. PLLA also experiences brittle failure in ambient conditions due to its T\textsubscript{g} of around 60\textdegree C, with strain at break of around 4\%, making deployment of PLLA-based medical devices (where some plastic deformation is often required) difficult \cite{Farah2016}.\\

In addition to its inherently low toughness, PLLA can also become embrittled over time, which can lead to catastrophic failure of implanted devices \cite{Pan2007, Wayangankar2015}. This embrittlement can be caused by either enthalpy relaxation or crystallisation. Enthalpy relaxation (also known as physical aging), is a result of the thermodynamically unstable nature of the fast-cooled glassy polymer - the material tends to equilibrium through slow rearrangements, resulting in densification without long-range order, greater intermolecular interactions, and loss of ductility \cite{Hodge1994, Pan2007}. Enthalpy relaxation of PLLA can occur after several days at temperatures below T\textsubscript{g} \cite{Pan2007}. Crystallisation is also driven by the non-equilibrium nature of the frozen glassy state, and requires higher chain mobility to form ordered structures. It is therefore typically seen when PLLA is heated above its T\textsubscript{g}, however it also occurs during degradation when hydration and hydrolysis-induced chain scission provide additional chain mobility \cite{Sarasua2005, Gleadall2012, Han2009}.\\

Polymer blending provides a convenient solution to tune the material properties of PLLA to address some of the drawbacks discussed above, however many of these blend systems (particularly  those including poly($\varepsilon$-caprolactone), a popular addition to improve ductility) are incompatible and form a phase separated, immiscible structure \cite{Broz2003, Lopez-Rodriguez2006}. To achieve better compatibility while also allowing tuning of polymer properties, blends of PLLA with poly(\textsc{L}-lactide-co-$\varepsilon$-caprolactone) (PLCL) have been developed and show improved miscibility \cite{Hiljanen-Vainio1996a, Ugartemendia2018,  Kang2016b}, however this promising blend system has not yet been extensively studied. Bioresorbable polymer properties including degradation rates, strength, and ductility must be enhanced beyond those of PLLA if we are to achieve key application requirements including lower strut thickness, sufficient radial strength, safe deployment performance, and reduced fracture risks needed to advance the technology.\\

PEGylation is a common technique for improving drug delivery by increasing solubility and enhancing biocompatibility \cite{Pasut2012}, however few studies have examined its effects when copolymerised with PLLA. There is evidence that addition of PEG (polyethylene glycol) to PLLA makes the polymer more hydrophilic, thus increasing the water absorption and degradation rate, as well as increasing the ductility \cite{Hu1993, Hu1994, Zhu1990a, Lee2005, Cohn1988}, however these all concern relatively small molecules with large PEG components. More recently PEG functionalised PLLA has been developed \cite{Azhari2018, Pacharra2018}, where the term functionalisation is used to emphasise the small quantity of PEG present in these polymers (0.1 - 1.5 wt\%). Despite the low PEG content, similar results are seen, with increased water absorption and degradation rate resulting from the presence (but not length) of PEG (up to PEG M\textsubscript{n} 5000 g mol\textsuperscript{-1}).\\

In this work, we aim to address some of the drawbacks of PLLA, by investigating blends of PLLA with PLCL-PEG (polyethylene glycol functionalised poly(\textsc{L}-lactide-co-$\varepsilon$-caprolactone)), allowing tuning of the mechanical and degradation properties of PLLA. We characterise the structure and mechanical properties of PLLA:PLCL-PEG blends, and examine how these change during degradation, revealing the key mechanisms involved in determining this behaviour.


\section{Materials and Methods}

\subsection{Materials}
PLLA (Ingeo 2500 HP) was supplied by Natureworks LLC, with M\textsubscript{w} of 184 kg mol\textsuperscript{-1} (measured after processing). PLCL-PEG was synthesised and supplied by Ashland Specialties Ireland Ltd. (Dublin, Ireland), with copolymer ratios of 80:20 and 70:30 (LA:CL), and a singular PEG end-group of length 550 g mol\textsuperscript{-1}. \textsuperscript{1}H-NMR was utilised to determine the chemical composition of the co-polymers and confirm the presence of PEG. The PLCL(80:20)-PEG and PLCL(70:30)-PEG had a M\textsubscript{w} of 188 kg mol\textsuperscript{-1} and 129 kg mol\textsuperscript{-1} respectively (measured after processing). Dichloromethane (DCM) was supplied by Merck KGaA, Germany, and Chloroform was purchased from Sigma Aldrich. Gibco phosphate-buffered saline (PBS), pH $=$ 7.4, was supplied by Thermo Fisher Scientific Inc., USA.

\subsection{Processing}
Polymer blends were prepared by blending PLLA with 10, 20, 30, 40, 50, and 60 wt\% PLCL(80:20)-PEG or PLCL(70:30)-PEG. Solvent cast films of pure and blended polymers were produced by dissolution in DCM (0.1 g mL\textsuperscript{-1}) and casting into petri dishes, before drying under vacuum at 50\textdegree C for 10 days until constant mass. Polymer films were then processed into dumbbell or disk shaped samples using micro-injection moulding (IM 5.5, Xplore Instruments BV, The Netherlands) and custom-made moulds in ambient conditions. Micro-injection moulding was carried out at the minimum melt temperature required for complete mould filling and uniform sample appearance, which ranged from 173 - 236\textdegree C depending on blend composition, with the mould held at ambient temperature. All samples were determined to be amorphous after moulding, with no sharp crystalline peaks observed in XRD patterns (see supplementary material).

\subsection{Characterisation}
Differential scanning calorimetry (DSC) was carried out using a DSC Q2000 (TA Instruments, USA), in hermetic Al pans at a heating rate of 20\textdegree C min\textsuperscript{-1}, from -20 to 230\textdegree C (-20 to 80\textdegree C for hydrated samples due to outgassing). TA Universal Analysis software was used to determine the glass transition temperature T\textsubscript{g}, taken at the inflection point. X-ray diffraction (XRD) was carried out using a Bruker D8 Advance diffractometer with Cu K$\alpha$ radiation in a 2$\theta$ range of 5-50\textdegree, with a 0.05\textdegree\ step size and dwell time of 1.0 s step\textsuperscript{-1}. The crystallinity was estimated using HighScore Plus (Malvern Panalytical), by fitting crystalline and broad amorphous peaks above the instrument background. Scanning electron microscopy (SEM) was done using an FEI Nova NanoSEM, using an accelerating voltage of 5 kV. Samples were prepared by cryo-fracturing in liquid nitrogen to view the cross-section, and then sputter coating with approximately 10 nm of gold (Emitech K550 Sputter Coater). Gel permeation chromatography (GPC) samples were prepared by dissolving 3 - 7 mg of polymer in 2 ml of chloroform followed by filtration using a Millipore filter (0.20 $\mu$m pore size, Dia 13mm, Millipore SLFG013NL, Fluropore PTFE (F) membrane). GPC measurements were performed using an Agilent triple detector system with an Agilent Technologies column (PLgel 5m MIXED-C 300x 7.5mm). The columns were calibrated with polystyrene standards supplied by Agilent Technologies at a concentration of 10 mg ml\textsuperscript{-1}. Chloroform (Sigma Aldrich) was used as the mobile phase with a flow rate of 1.0 ml min\textsuperscript{-1}. Molecular weight data was collected using the refractive index peak.

\subsection{Mechanical testing}
Tensile testing was carried out using an H5KS Benchtop Tester (Tinius Olsen Ltd, UK) with a 250 N load cell, under a constant elongation rate of 2 mm min\textsuperscript{-1}. Dumbbell samples (5 mm gauge length) were tested in ambient (dry at 25\textdegree C) and simulated body conditions (immersed in deionised water at 37\textdegree C) using a Saline Test Tank with Heater (Tinius Olsen Ltd, UK). After loading samples into the grips and immersing them in water, they were left for approximately 10 minutes for the temperature to equilibrate. Strain was measured using a video extensometer and custom-built LabVIEW software. Yield strength ($\sigma_{y}$) for polymers was taken as the 0.2\% offset yield point, and the elastic modulus (E) was calculated from the linear region of the stress-strain curve before yield.

\subsection{Degradation study}
Degradation studies were carried out by immersing individual disk-shaped polymer samples in 5 mL PBS in bijou tubes, which were placed in an incubator at 37\textdegree C. pH measurements were taken at regular intervals using an HI 4222 pH meter (Hanna Instruments Ltd., UK), and PBS alone was used as a control for pH measurements. A long-term (480 days) degradation study was carried out, as well as a shorter (30 days) study for more detailed analysis.


\section{Results}

\subsection{Initial structure}

\begin{figure}
	\centering
	\includegraphics[width=0.7\linewidth]{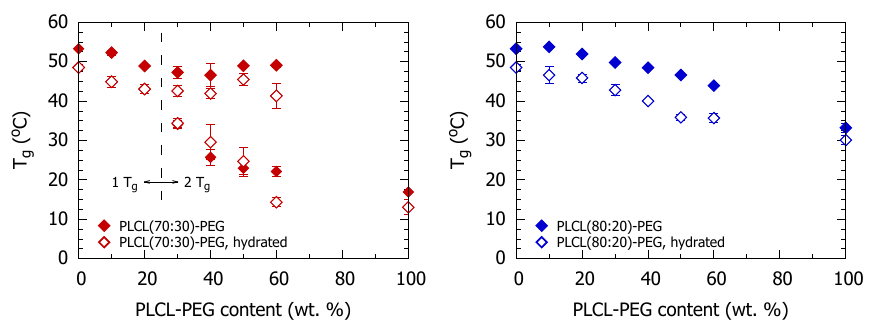}
	\caption{Glass transition temperature(s) measured by DSC for different PLLA:PLCL-PEG blend compositions, in dry and hydrated (for 20 minutes in deionised water) states. Blends of PLLA with PLCL(70:30)-PEG (left) and PLCL(80:20)-PEG (right) are shown. Error bars denote standard deviation, n = 3.}
	\label{fig:DSC_t0}
\end{figure}

Measured glass transitions of injection-moulded polymer blends are shown in Fig. \ref{fig:DSC_t0}, and it is clear that blends of PLLA with PLCL(80:20)-PEG were completely miscible across the composition range studied, showing a single T\textsubscript{g} for each composition. Blends of PLLA with PLCL(70:30)-PEG were partially miscible, with blends containing $\leq$ 20 wt\% of the copolymer showing single phase behaviour, and blends with $\geq$ 30 wt\% were phase separated. The effect of hydration (for 20 minutes in deionised water) can also be seen - there was a general decrease in the T\textsubscript{g} of approximately 5\textdegree C.\\

\begin{figure*}
	\centering
	\includegraphics[width=0.8\linewidth]{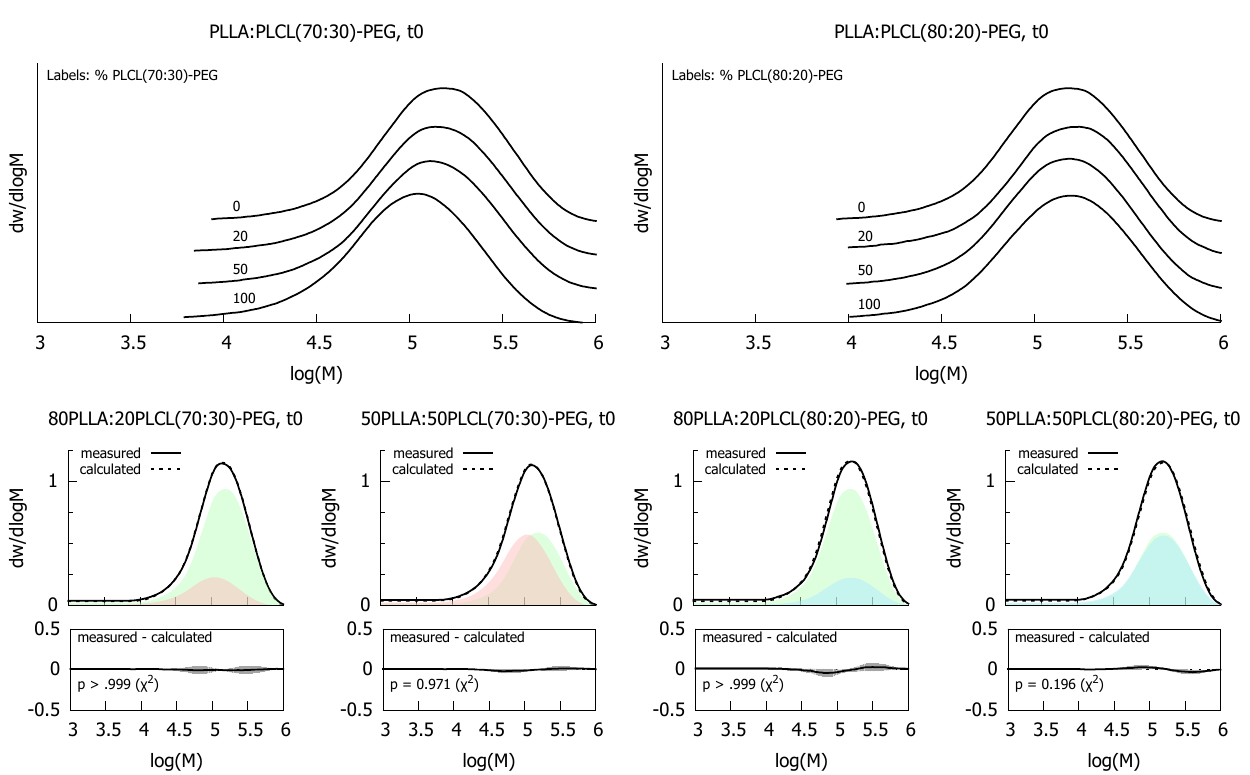}
	\caption{GPC molecular weight distributions for as-fabricated PLLA:PLCL(70:30)-PEG blends (top left) and PLLA:PLCL(80:20)-PEG blends (top right); curves are offset for clarity. Bottom row shows measured and calculated distributions, based on a linear combination of the individual components (green = PLLA, red = PLCL(70:30)-PEG, blue = PLCL(80:20)-PEG). Shaded region in difference plot denotes standard deviation, n = 3.}
	\label{fig:GPC_t0}
\end{figure*}

From the molecular weight distributions (Table \ref{tab:GPC}, Fig. \ref{fig:GPC_t0}), we see that all the pure polymers had comparable molecular weights, with 184 ($\pm$ 8) kg mol\textsuperscript{-1} for PLLA, 188 ($\pm$ 4) kg mol\textsuperscript{-1} for PLCL(80:20)-PEG, and PLCL(70:30)-PEG slightly lower at 129 ($\pm$ 4) kg mol\textsuperscript{-1}. Theoretical distributions for the polymer blends were calculated based on a linear combination of the two component polymer distributions, and these calculated distributions compared well with the measured distributions. This is indicated by the small difference between measured and calculated distributions, and the associated p-value, indicating no evidence of a difference between the distributions (chi-squared test $p > 0.05$, $H_{0}$ = no difference between distributions).

\begin{table*}
	\centering
	\caption{\label{tab:GPC}{Molecular weights of PLLA:PLCL-PEG blends before and after 30 days degradation, along with calculated degradation rate according to $ln\:M_{n}(t_{2}) = ln\:M_{n}(t_{1}) - kt$. Error shown is the standard deviation from three measurements.}}
	{\renewcommand{\arraystretch}{1.2}
		\begin{tabularx}{\linewidth}{ X X X X X X X}
			\toprule
			& \multicolumn{2}{c}{\textit{\textbf{M\textsubscript{w},t0}} (kg mol\textsuperscript{-1})} & \multicolumn{2}{c}{\textit{\textbf{M\textsubscript{w},t30}} (kg mol\textsuperscript{-1})} & \multicolumn{2}{c}{\textit{\textbf{k}} (10\textsuperscript{-3} day\textsuperscript{-1})} \\
			wt\% PLCL-PEG & PLCL(80:20)-PEG & PLCL(70:30)-PEG & PLCL(80:20)-PEG & PLCL(70:30)-PEG & PLCL(80:20)-PEG & PLCL(70:30)-PEG \\
			\midrule
			0 & \multicolumn{2}{c}{184 ($\pm8$)} & \multicolumn{2}{c}{156 ($\pm2$)} & \multicolumn{2}{c}{7.9 ($\pm1.9$)}  \\
			20 & 190 ($\pm5$) & 173 ($\pm6$) & 121 ($\pm4$) & 85 ($\pm3$) & 16.7 ($\pm3.0$) & 28.2 ($\pm1.8$)  \\
			50 & 181 ($\pm2$) & 159 ($\pm2$) & 62 ($\pm2$) & 58 ($\pm1$) & 41.9 ($\pm0.9$) & 54.7 ($\pm1.3$)  \\
			100 & 188 ($\pm4$) & 129 ($\pm4$) & 53 ($\pm2$) & 27.7 ($\pm0.6$) & 48.6 ($\pm1.9$) & 58.7 ($\pm1.4$)  \\
			\bottomrule 
		\end{tabularx}
	}
\end{table*}

\subsection{Initial mechanical properties}

\begin{figure}
	\centering
	\includegraphics[width=0.6\linewidth]{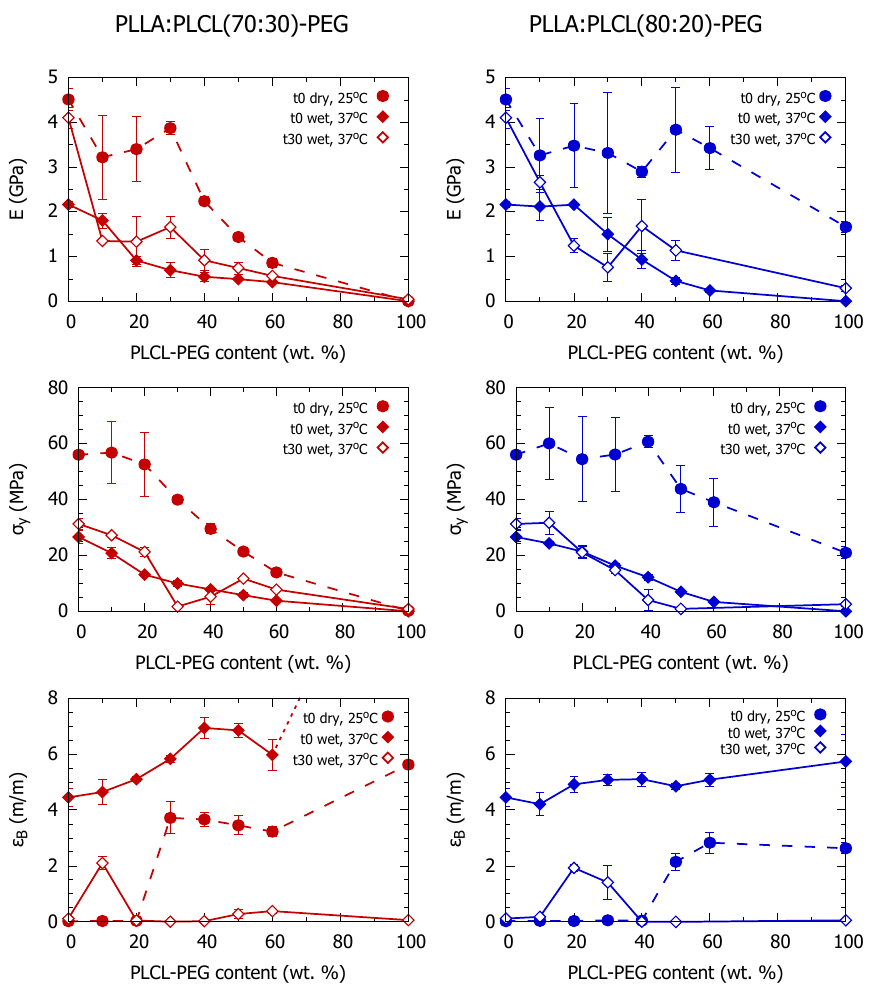}
	\caption{Mechanical properties measured by tensile testing for PLLA:PLCL-PEG blends under various conditions. \textbf{t0 dry, 25\textdegree C} denotes as-fabricated samples tested dry at room temperature,  \textbf{t0 wet, 37\textdegree C} denotes as-fabricated samples tested immersed in 37\textdegree C water, and \textbf{t30 wet, 37\textdegree C} denotes samples tested immersed in 37\textdegree C water, after 30 days degradation in PBS. Top row: elastic modulus, middle row: yield strength, bottom row: elongation at break. Left column: PLLA:PLCL(70:30)-PEG blends, right column: PLLA:PLCL(80:20)-PEG blends. Error bars denote standard deviation, n = 3.}
	\label{fig:Mech_Summ}
\end{figure}

Under ambient conditions (dry at 25\textdegree C), the addition of the softer PLCL-PEG to PLLA reduced its strength and stiffness (Fig. \ref{fig:Mech_Summ}) as has been reported previously for PLLA:PLCL blends \cite{Hiljanen-Vainio1996a, Ugartemendia2018}, with the PLCL(70:30)-PEG copolymer having a larger effect than PLCL(80:20)-PEG. Pure PLLA was brittle as expected ($\varepsilon_{B} \sim 2.4 \%$), as were blends with small amounts of PLCL-PEG. However, once a certain amount of copolymer was added ($\geq$ 30 wt\% PLCL(70:30)-PEG, or $\geq$ 50 wt\% PLCL(80:20)-PEG), a step change was seen and large scale ductility was observed ($\varepsilon_{B} \sim$ 200 - 400\%). This step change occurred at the same composition as the formation of a PLCL-PEG-rich phase in the polymer blend, either by phase separation in the case of PLLA:PLCL(70:30)-PEG blends (Fig. \ref{fig:DSC_t0}), or by simply altering the bulk composition in the case of miscible PLLA:PLCL(80:20)-PEG blends.\\

Under simulated body conditions (immersed in deionised water at 37\textdegree C), the combined effect of hydration and elevated temperature had a dramatic effect, reducing the strength and stiffness while increasing the ductility such that all polymers tested, including pure PLLA, exhibited ductile failure. Pure PLCL(70:30)-PEG, which showed the greatest elongation at break of 560\% in ambient conditions,  was deformed to 1600\% and reached the maximum travel of the water bath without fracture (denoted by dotted line in Fig. \ref{fig:Mech_Summ}).\\

\subsection{Long-term degradation}

\begin{figure}
	\centering
	\includegraphics[width=0.7\linewidth]{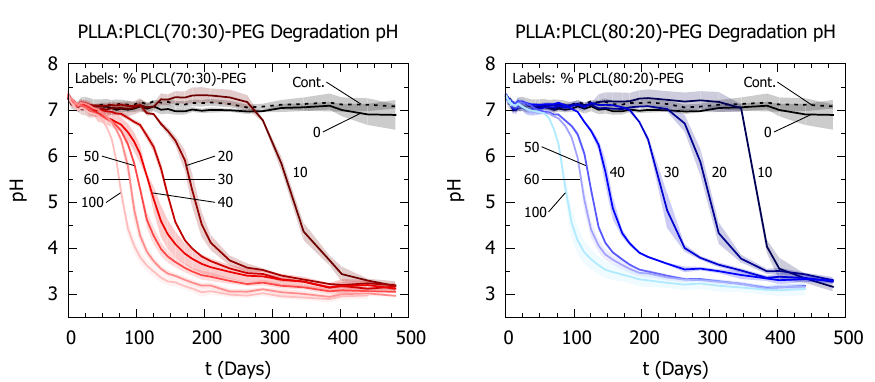}
	\caption{Solution pH during long-term degradation test of PLLA:PLCL-PEG blends in PBS at 37\textdegree C. Shaded region denotes standard deviation, n = 3.}
	\label{fig:Deg_pH}
\end{figure}

\begin{figure}
	\centering
	\includegraphics[width=0.7\linewidth]{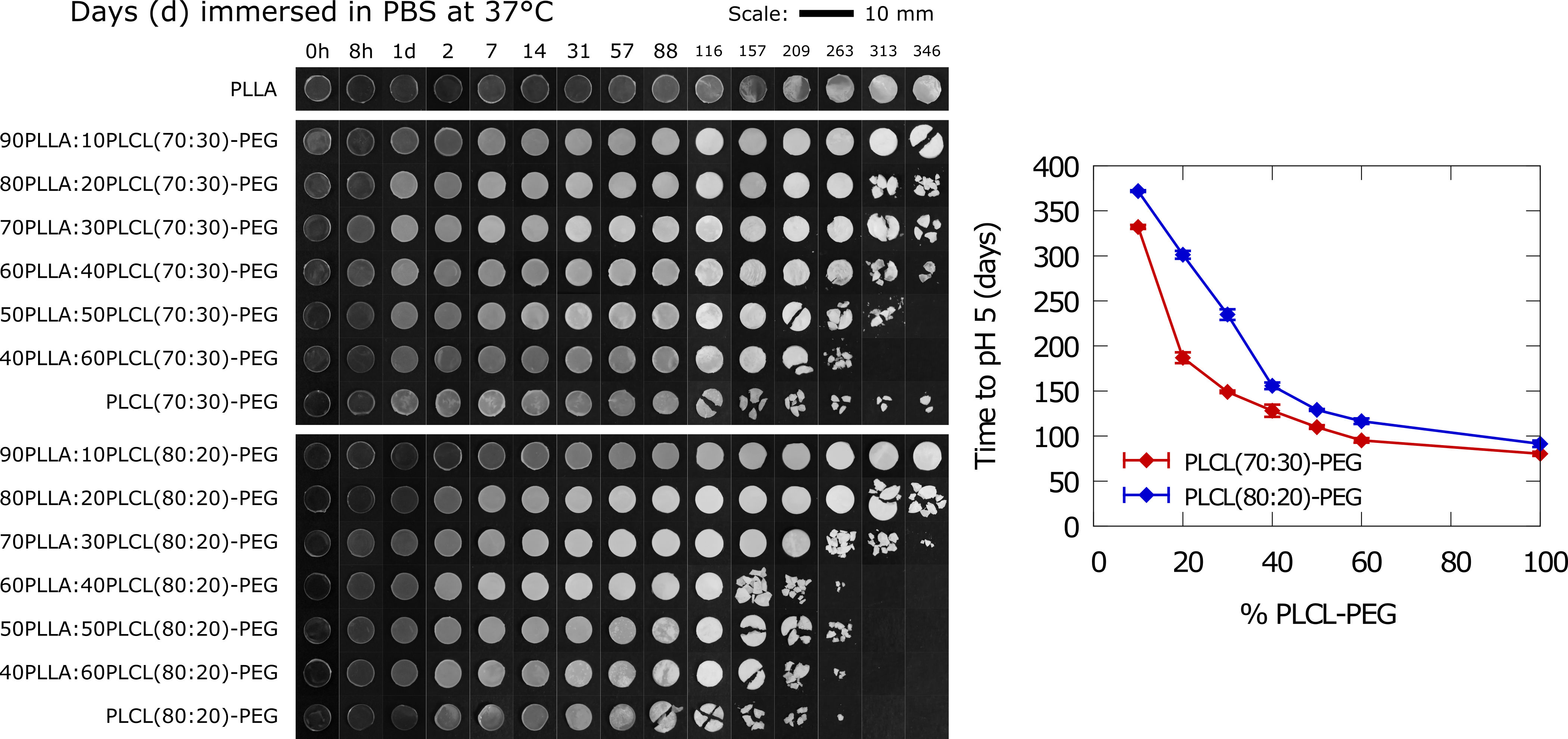}
	\caption{Left: photographs showing polymer blend disk morphology during degradation. Right: time taken to reach a solution pH = 5 during degradation, giving an indication of the relative degradation rate.}
	\label{fig:Morph_fig}
\end{figure}

During long-term immersion degradation tests (Fig. \ref{fig:Deg_pH}, \ref{fig:Morph_fig}), degradation of the polymer into lactic acid, causing pH reduction, was seen for all polymer and blend compositions except pure PLLA, which did not show any pH changes during the timescale of this test. Pure PLCL-PEG polymers showed rapid degradation, resulting in pH reduction after approximately 2 months. The PLLA:PLCL-PEG blends all showed intermediate degradation behaviour, with the addition of increasing amounts of PLCL-PEG accelerating degradation, demonstrating the ability to controllably accelerate PLLA degradation via blending with PLCL-PEG. PLCL(70:30)-PEG degraded faster than PLCL(80:20)-PEG due to the higher CL content, and therefore also accelerated the blend degradation to a greater extent. The degradation time showed the same trend, and makes clear that the dependence on composition is not linear (Fig. \ref{fig:Morph_fig}), with small initial PLCL-PEG additions causing a large increase in degradation rate.\\

Examining the polymer morphology (Fig. \ref{fig:Morph_fig}) can also reveal information about the degradation behaviour. Fragmentation of the polymers was seen at advanced stages during degradation, and the onset of fragmentation roughly corresponded to the speed of degradation. Fragmentation appeared to begin slightly earlier for PLLA:PLCL(80:20)-PEG blends than for PLLA:PLCL(70:30)-PEG, even though the former degraded more slowly, which may be a result of the more brittle nature of the higher LA copolymer. All the polymers were also seen to turn from transparent to opaque during degradation, which could indicate crystallisation or other processes resulting in light scattering.\\

\subsection{Structure after short-term degradation}

\begin{figure*}
	\centering
	\includegraphics[width=0.8\linewidth]{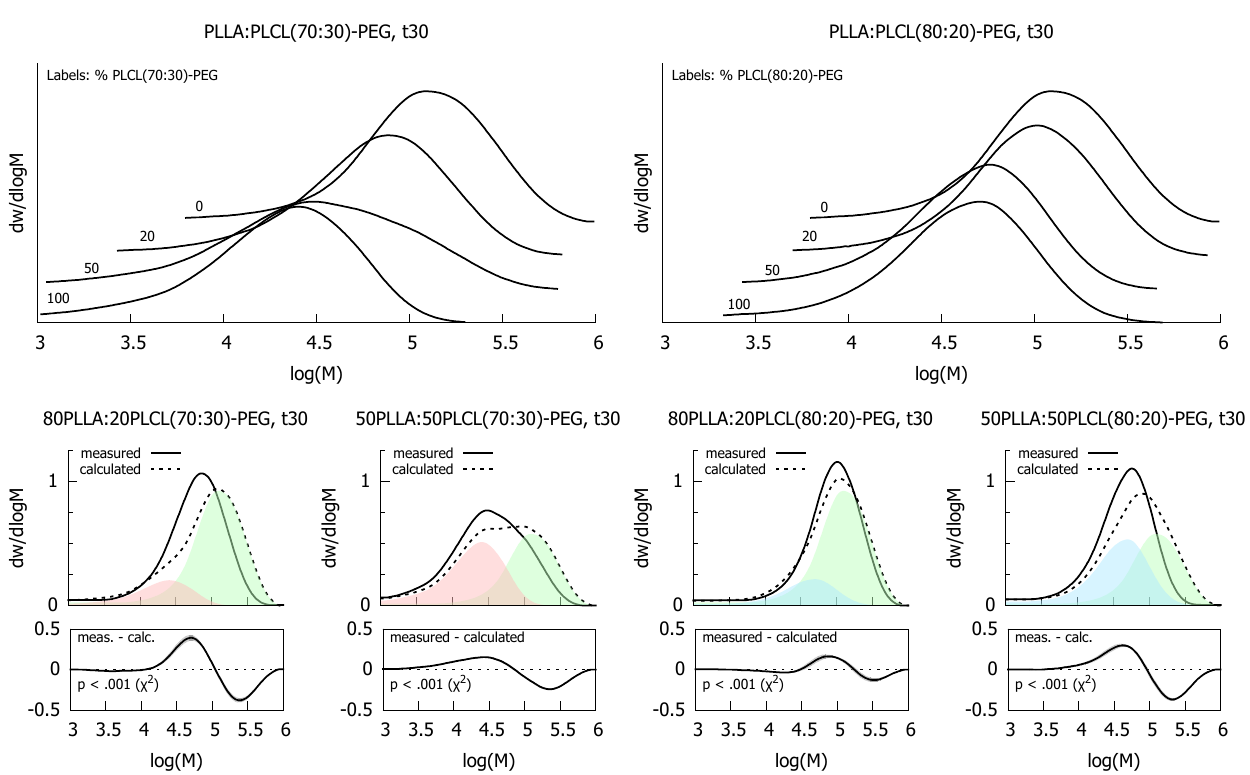}
	\caption{GPC molecular weight distributions for PLLA:PLCL(70:30)-PEG blends (top left) and PLLA:PLCL(80:20)-PEG blends (top right) after 30 days degradation in PBS; curves are offset for clarity. Bottom row shows measured and calculated distributions, based on a linear combination of the individual components (green = PLLA, red = PLCL(70:30)-PEG, blue = PLCL(80:20)-PEG). Shaded region in difference plot denotes standard deviation, n = 3.}
	\label{fig:GPC_t30}
\end{figure*}

\begin{figure}
	\centering
	\includegraphics[width=0.7\linewidth]{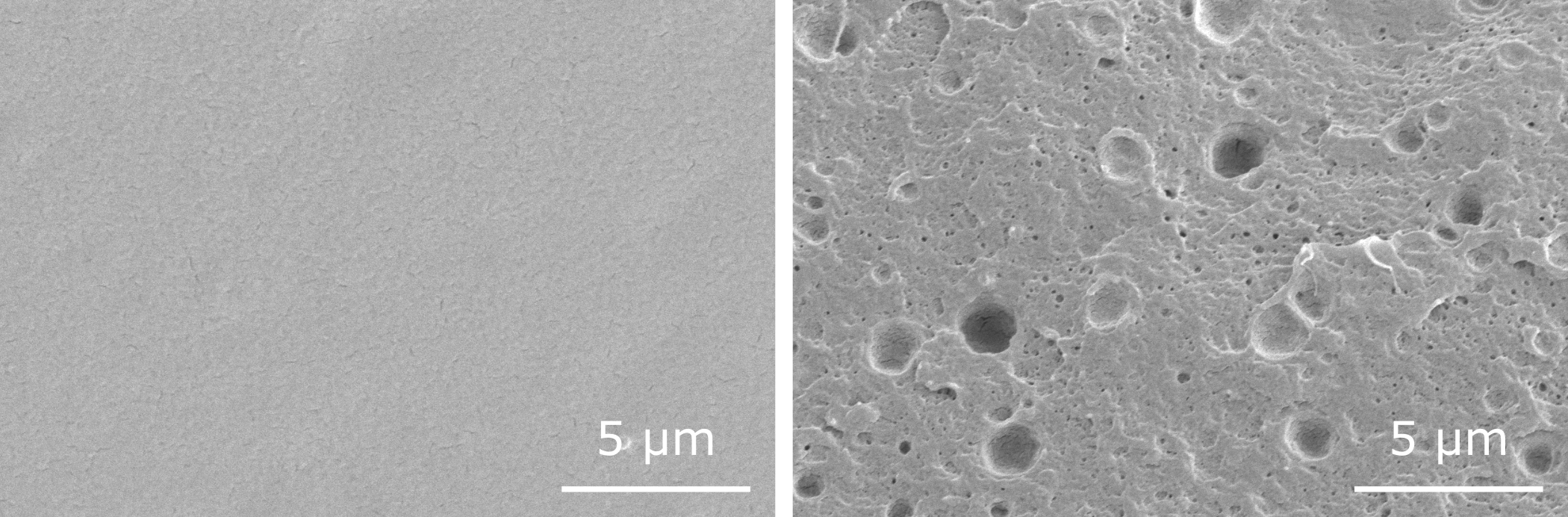}
	\caption{SEM images of cross-sections of 80PLLA:20PLCL(70:30)-PEG before (left) and after (right) 30 days degradation, showing voids appearing after degradation.}
	\label{fig:SEM_t30}
\end{figure}

After 30 days degradation, GPC showed varying levels of molecular weight reduction, as seen in Table \ref{tab:GPC} and Fig. \ref{fig:GPC_t30}. Pure PLLA displayed a small amount of degradation, with a degradation rate comparable with previous works \cite{Tsuji2003a, Sin2012}, while the addition of PLCL-PEG copolymers to the blend significantly increased the degradation rate. As expected, this increase was greater for addition of the less stable PLCL(70:30)-PEG copolymer, consistent with similar results seen by pH measurement during long term degradation. Examining the molecular weight distributions (Fig. \ref{fig:GPC_t30}), it is clear that there was a significant difference between the measured distribution and those calculated from individually degrading components. This is shown by the large difference plot and associated p-value, indicating strong evidence of a difference between the distributions (chi-squared test $p < 0.05$, $H_{1}$ = difference between distributions exists). In all cases, after degradation the measured molecular weight distribution showed a lower amount of high molecular weight components than would be expected from degradation of PLLA, indicating that not only did the PLCL-PEG component degrade, but that this also caused accelerated degradation of the PLLA component when compared with how PLLA degraded on its own. When examining polymer blends by SEM before and after degradation, the appearance of voids within the structure was observed (Fig. \ref{fig:SEM_t30}), although care must be taken in interpreting such images because of the potential effect of dehydration before imaging on the microstructure.\\

\begin{figure}
	\centering
	\includegraphics[width=0.7\linewidth]{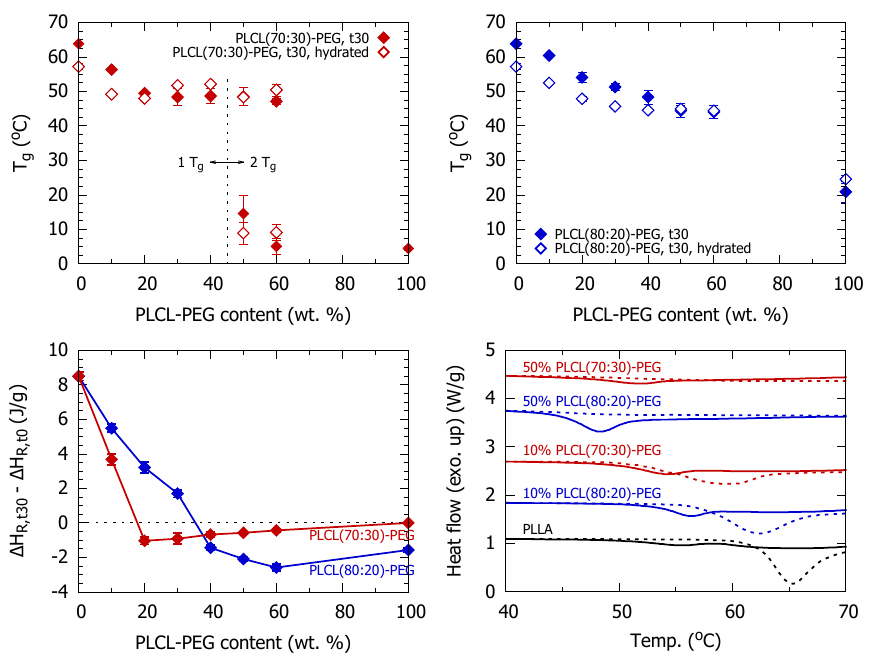}
	\caption{Thermal properties of PLLA:PLCL-PEG blends measured after 30 days degradation in PBS at 37\textdegree C. Glass transition temperature (T\textsubscript{g}) of PLLA:PLCL(70:30)-PEG blends (top left) and PLLA:PLCL(80:20)-PEG blends (top right) in dry and hydrated conditions. The change in enthalpy relaxation peak area during degradation (bottom left), and examples of the glass transition region of DSC curves (bottom right) are also shown. For DSC curves, solid lines denote measurements before degradation, and dotted lines denote measurements after 30 days degradation. Error bars denote standard deviation, n = 3.}
	\label{fig:DSC_t30}
\end{figure}

DSC analysis (Fig. \ref{fig:DSC_t30}) revealed several changes to the polymer blend structure that had occurred during 30 days degradation in PBS. While the PLLA:PLCL(80:20)-PEG blends showed unchanged miscibility behaviour, that of the PLLA:PLCL(70:30)-PEG blends had changed, with an increased region of miscibility now up to $\leq$ 40 wt\% PLCL(70:30)-PEG. Changes were also seen in the enthalpy relaxation behaviour, where the associated peak (endothermic peak after the glass transition - $\Delta H_{R}$) significantly increased after degradation for pure PLLA and blends with low PLCL-PEG content (Fig. \ref{fig:DSC_t30}). Examples of the glass transition region of the DSC curve are also shown in Fig \ref{fig:DSC_t30}, where the increase in the endothermic enthalpy relaxation peak after the glass transition can be seen for pure PLLA and 10 wt\% PLCL-PEG blends, but not for blends with higher PLCL-PEG content. In addition to the increased endothermic peak after the glass transition, blends that displayed significant enthalpy relaxation also showed increased glass transition temperatures compared to their as-fabricated state.\\

\begin{figure}
	\centering
	\includegraphics[width=0.7\linewidth]{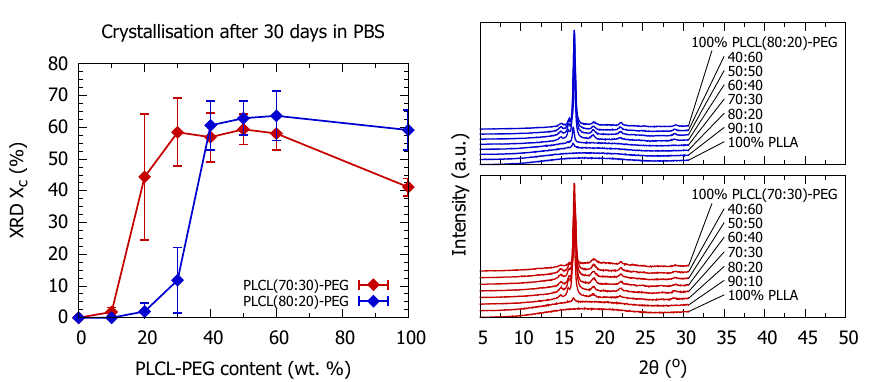}
	\caption{Left: XRD crystallinity of initially amorphous polymer blends after 30 days degradation in PBS. Error bars denote standard deviation, n = 3. Right: representative diffraction patterns for PLLA:PLCL-PEG blends after degradation, labels denote PLLA:PLCL-PEG blend ratio, and datasets are offset for clarity.}
	\label{fig:XRD_t30}
\end{figure}

XRD was used to determine whether the initially amorphous polymer blends had crystallised during degradation. The crystalline content and representative diffraction patterns are shown in Fig. \ref{fig:XRD_t30}. For pure PLLA and blends with low PLCL-PEG content, little to no crystallisation was seen. At higher PLCL-PEG content a step change was seen, where a large increase in crystallinity took place, increasing to a plateau at around 60\% crystallinity. Observed peaks are consistent with the $\alpha$ form of the PLLA unit cell \cite{Wasanasuk2011, Tashiro2017}.\\

\subsection{Mechanical properties after short-term degradation}\label{subsec:Mech_t30}

After 30 days degradation in PBS at 37\textdegree C, several significant changes were seen in the mechanical properties of the polymer blends compared with those measured under the same conditions (in 37\textdegree C water) before degradation (Fig. \ref{fig:Mech_Summ}). Firstly, pure PLLA and 90PLLA:10PLCL(80:20)-PEG were been strengthened but embrittled, with an increase in stiffness and strength, and large reduction in ductility. Blends with $\geq$ 20\% PLCL(70:30)-PEG or $\geq$ 40\% PLCL(80:20)-PEG also displayed significant embrittlement. Finally, the group of blends of PLLA with 10\% PLCL(70:30)-PEG or 20 - 30\% PLCL(80:20)-PEG constitute a compositional ``sweet spot'', where the polymers retained a significant amount of ductility after 30 days degradation.


\section{Discussion}

The initial phase separation behaviour of PLLA:PLCL(70:30)-PEG blends observed here (Fig. \ref{fig:DSC_t0}) is comparable with that reported by Ugartemendia \textit{et al.} \cite{Ugartemendia2018}, who observed phase separation at $\geq$ 40 wt\% PLCL(67:33) (for melt-blending) - the small difference observed could be attributed to the different thermal history, or an effect of PEG functionalisation. The subsequent reduction in T\textsubscript{g} upon hydration can be attributed to water absorption and plasticisation, as water molecules diffuse between polymer chains, increasing free volume and reducing inter-chain bonding \cite{Kim2004, Damico2014, Vyavahare2014, Renouf-Glauser2005}. This has a profound impact on the mechanical properties, resulting in the increased ductility observed here (Fig. \ref{fig:Mech_Summ}), consistent with recent works illustrating the plasticising effect of water on the bulk mechanical properties of PLLA \cite{Dreher2017, Wang2018}.\\

During degradation, no pH reduction is seen for pure PLLA. This is as expected as it is known to require over a year for resorption \cite{Mukherjee2017, Serruys1988, Maurus2004, McMahon2018, Farah2016}. PLCL copolymers have been shown to degrade faster than either homopolymer \cite{Malin1996}, and this rapid degradation is seen here after approximately 2 months. After 30 days  GPC measurements of degradation indicate that the addition of PLCL-PEG to PLLA accelerates degradation of the polymer blend. This can be attributed to auto-catalytic behaviour \cite{Gleadall2014, Tsuji2002} - as the faster degrading PLCL-PEG breaks down, there is an accumulation of catalytic oligomers formed by hydrolysis, which then catalyse degradation of PLLA as well as further degradation of PLCL-PEG. This is consistent with SEM observations, where the voids seen may correspond to areas where significant auto-catalysis has occurred, resulting in pockets of low molecular weight oligomers which diffuse slowly out through the polymer matrix \cite{Gleadall2012, Gleadall2014}. The non-linear dependence of the degradation time on composition (Fig. \ref{fig:Morph_fig}) also corroborates this autocatalytic mechanism.\\

Structural changes in polymer blends after degradation include increased miscibility, enthalpy relaxation, and crystallisation. The increased miscibility of PLLA:PLCL(70:30)-PEG blends can be attributed to the molecular weight reduction that takes place during degradation, making mixing more thermodynamically favourable. The increase in the enthalpy relaxation peak ($\Delta H_{R}$) for pure PLLA and blends with low PLCL-PEG content indicates that these blends have undergone rearrangement into a denser, more thermodynamically stable configuration. This has also resulted in an increased T\textsubscript{g} due to the reduced free volume in the densified structure \cite{Pan2007}. The chain cleavage resulting from the hydrolysis reaction provides additional mobility for the polymer to rearrange into a crystalline structure \cite{Han2009}, leading to those blends that degrade faster (greater PLCL-PEG content, and greater CL content in copolymer) having higher crystalline content. After the step transition ($\geq$ 20\% PLCL(70:30)-PEG or $\geq$ 40\% PLCL(80:20)-PEG) the blends show a similar degree of crystallinity, indicating that some maximum extent of crystallisation has been reached. Pure PLCL(70:30)-PEG shows a slightly lower crystallinity however, probably due to the difficulty of packing the less homogeneous copolymer. The appearance of the polymer disks during the long term degradation study (Fig. \ref{fig:Morph_fig}) is also consistent with crystallisation, where even the blends with low crystalline content show some opacity, while pure PLLA which has not crystallised after 30 days degradation remains transparent. It must be acknowledged however, that opacity could also arise as a result of cracking that may occur during degradation.\\

These structural changes have a direct result on the mechanical properties. Pure PLLA and 90PLLA:\ 10PLCL(80:20)-PEG samples experience significant enthalpy relaxation, leading to densification, increased stiffness and strength, but lower ductility \cite{Pan2007}. Blends with $\geq$ 20\% PLCL(70:30)-PEG or $\geq$ 40\% PLCL(80:20)-PEG display reduced molecular weight and high crystallinity as described above, resulting in increased stiffness and variable gains in yield strength, along with severe brittleness. The presence of a small amount of PLCL-PEG in blends of PLLA with 10\% PLCL(70:30)-PEG or 20 - 30\% PLCL(80:20)-PEG reduces the enthalpy relaxation during this timeframe (Fig. \ref{fig:DSC_t30}), due to the difficulty of structural rearrangement when two different polymers are present. Also, the amount of PLCL-PEG present is not sufficient to cause significant chain cleavage which leads to crystallisation (Fig. \ref{fig:XRD_t30}), resulting in a structure that is more resistant to enthalpy relaxation and crystallisation, therefore retaining much of its original ductility.\\

\begin{figure}
	\centering
	\includegraphics[width=0.5\linewidth]{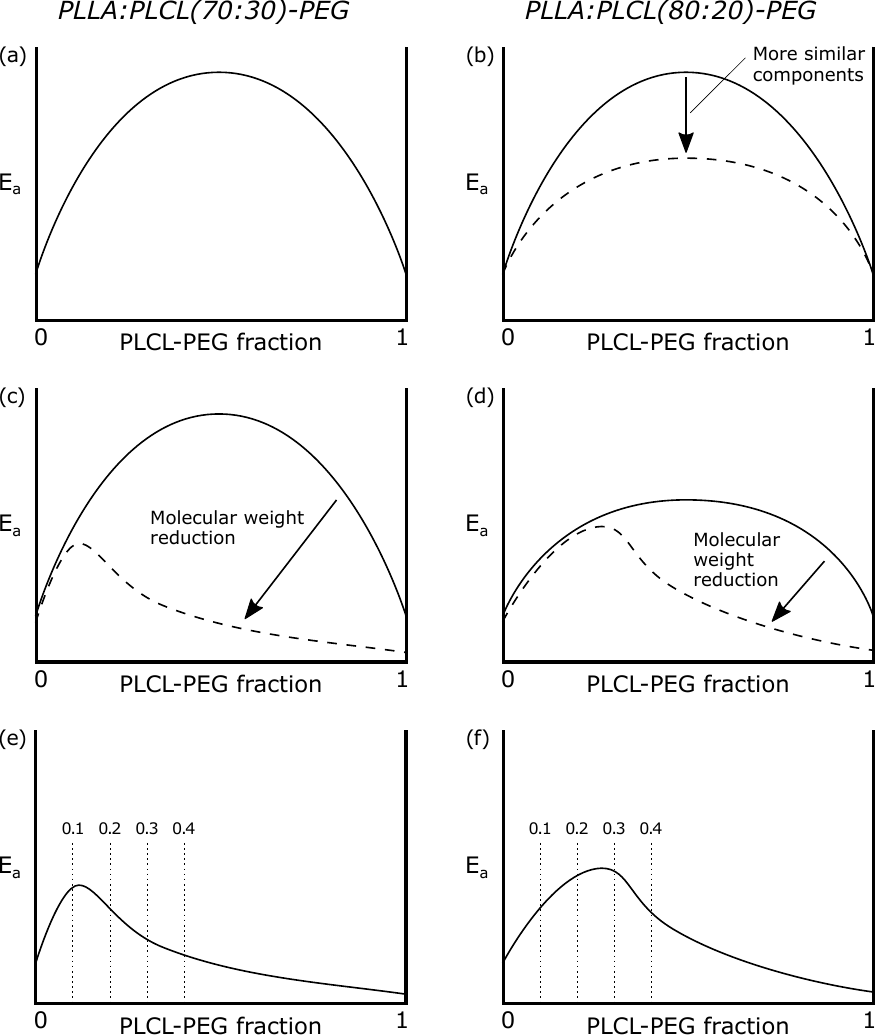}
	\caption{Schematic diagrams showing the activation energy (E\textsubscript{a}) for the general structural changes across blend compositions, showing the effect of polymer composition (a, b) and molecular weight reduction (c, d). (e, f) shows the combined effect of polymer composition and degradation, resulting in a maximum activation energy at certain compositions.}
	\label{fig:Ea_schematic}
\end{figure}

Based on the results discussed above, we suggest the following theoretical framework to summarise and explain the evolution of mechanical properties during degradation, specifically the embrittlement that occurs during degradation, resulting in the initially ductile polymers (when tested in 37\textdegree C water) becoming brittle. This embrittlement has been shown to be the result of two different structural changes, which occur for different blend ratios: enthalpy relaxation, and crystallisation. These are both exothermic transformations which have an associated activation energy, which governs the rate of transformation \cite{Pan2007, Vasanthakumari1983}. In general for polymer blends, like-like interactions are favoured over interactions between different polymers, i.e. the Flory-Huggins interaction parameter $\chi$ is generally positive. It then follows that the activation energy for the general structural relaxation which increases intermolecular interactions will be greatest for a 50:50 blend, and lowest for the pure polymers. This is shown schematically in Fig.  \ref{fig:Ea_schematic}.a for PLLA:PLCL-PEG blends. By the same reasoning, if one of the blend components is exchanged for one that is chemically more similar to the other (e.g. exchanging PLCL(70:30)-PEG for PLCL(80:20)-PEG, which is more similar to PLLA), the magnitude of the activation energy will be decreased overall, as seen in Fig. \ref{fig:Ea_schematic}.b. \\

The activation energy for structural relaxation is also dependent on the molecular weight of a polymer, with a lower molecular weight known to reduce the activation energy and accelerate both enthalpy relaxation and crystallisation \cite{Pan2007, Vasanthakumari1983, Andreozzi2005}. This is relevant to this work due to the effect of degradation, which reduces the molecular weight. This molecular weight reduction is asymmetric however, as discussed above, with high PLCL-PEG blends degrading significantly more than blends with lower PLCL-PEG content. This is reflected in Fig. \ref{fig:Ea_schematic}.c-d, where the activation energy for structural relaxation is reduced significantly for high PLCL-PEG blends, but only slightly for blends with low PLCL-PEG content.\\

The combination of these effects, as summarised in Fig. \ref{fig:Ea_schematic}.e-f, results in a maximum activation energy for low PLCL-PEG content. With increasing PLCL-PEG, the activation energy increases due to interactions between dissimilar polymer chains. As we increase the PLCL-PEG content further however, the activation energy decreases again due to molecular weight reduction caused by the accelerating effects of degrading PLCL-PEG. This can therefore provide an explanation of the changes in mechanical properties seen in Fig. \ref{fig:Mech_Summ}, where a ``sweet spot'' of low PLCL-PEG content is seen, which retains some of its ductility during degradation. Higher or lower concentrations of PLCL-PEG result in stiffer but more brittle materials due to the aforementioned structural changes.\\

This theoretical framework can provide a useful approach to explaining the evolution of the mechanical properties of polymer blends during degradation, as well as providing a basis for rational design of blends to engineer desired properties.\\

When considering the design of materials for bioresorbable stents, these materials show significant promise. The double T\textsubscript{g} behaviour of the phase separated blends significantly increases the ductility of the dry polymer, allowing plastic deformation during the crimping process, while the hydration-induced plasticisation would allow expansion 
\textit{in vivo} with reduced risk of fracture. The accelerated degradation achieved here compared with pure PLLA is a promising result, allowing faster resorption on timescales that more closely resemble healing times, and the ability to prevent embrittlement during degradation is another crucial benefit. These results together provide a number of methods for designing  resorbable stent materials with improved mechanical properties, faster resorption time, and retention of favourable mechanical properties during degradation.

\section{Conclusions}
A blending strategy has been developed which facilitated fine-tuning of performance with respect to polymer ductility, strength and stiffness while also offering broadened flexibility and perspective to bioresorbable polymer degradation profiles.\\

The structure and mechanical properties of a set of PLLA:PLCL-PEG blends were characterised, both in their as-fabricated state and after 30 days degradation. In the dry state, low PLCL-PEG blends were brittle, however once a PLCL-PEG-rich phase was formed, blends became ductile. In simulated body conditions (in 37\textdegree C water) the effect of hydration plasticises the system reduced interchain bonding, resulting in reduced strength and stiffness, and significant ductility for all blends.\\

Degradation tests indicated that the presence of PLCL-PEG in the blend accelerates degradation; not only did the PLCL-PEG phase degrade, but the degradation products released accelerated degradation of the both components via auto-catalysis.\\

The evolution of the mechanical properties during degradation is governed by the structural changes that take place. Pure PLLA and blends with 10\% PLCL(80:20)-PEG display densification via enthalpy relaxation, leading to increased strength and stiffness, along with embrittlement. Slightly higher amounts of PLCL-PEG (10\% PLCL(70:30)-PEG, or 20-30\% PLCL(80:20)-PEG) reduce this tendency for enthalpy relaxation, decreasing the extent of embrittlement that takes place during degradation. Larger amounts of PLCL-PEG ($\geq$ 20\% PLCL(70:30)-PEG, or $\geq$ 40\% PLCL(80:20)-PEG) begin to cause extensive molecular weight reduction, enhancing chain mobility and allowing crystallisation. This also leads to increased strength and stiffness, along with embrittlement. These mechanisms, and the resulting compositional ``sweet spot'' have been explained in terms of an activation energy for structural relaxation, where the chemical similarity of the blend components, and changes in molecular weight, are the main controlling factors.

\section*{Acknowledgements}
The authors thank Lucideon Ltd. for financial support of the project, and Ashland Specialties Ireland Ltd. for providing materials. RNO would also like to thank the Woolf Fisher Trust and the Cambridge Trust, for provision of a PhD scholarship. The authors would also like to thank Mr Wayne Skelton-Hough, Mr Andrew Rayment, and Mr Robert Cornell for their technical support and helpful discussions throughout the project. Original data for this paper can be found at \url{https://doi.org/10.17863/CAM.41718}.

\bibliographystyle{elsarticle-num}

\bibliography{Refs}

\FloatBarrier

\clearpage
\newpage

\section*{Supplementary Material: Tuning structural relaxations, mechanical properties, and degradation timescale of PLLA during hydrolytic degradation by blending with PLCL-PEG}
\begin{flushleft}
	Reece N. Oosterbeek, Kyung-Ah Kwon, Patrick Duffy, Sean McMahon, Xiang C. Zhang, Serena M. Best, Ruth E. Cameron
	\bigskip
\end{flushleft}

\FloatBarrier

After injection moulding polymers were examined by XRD to determine whether crystallisation had taken place during processing. All samples were determined to be amorphous after injection moulding, with no crystalline peaks observed in XRD patterns (Fig. \ref{fig:XRD_t0}). This can be attributed to the fast cooling provided by the room temperature mould, quenching the polymer quickly into an amorphous state.

\renewcommand{\thefigure}{S\arabic{figure}}
\setcounter{figure}{0}
\begin{figure}[h]
	\centering
	\includegraphics[width=0.7\linewidth]{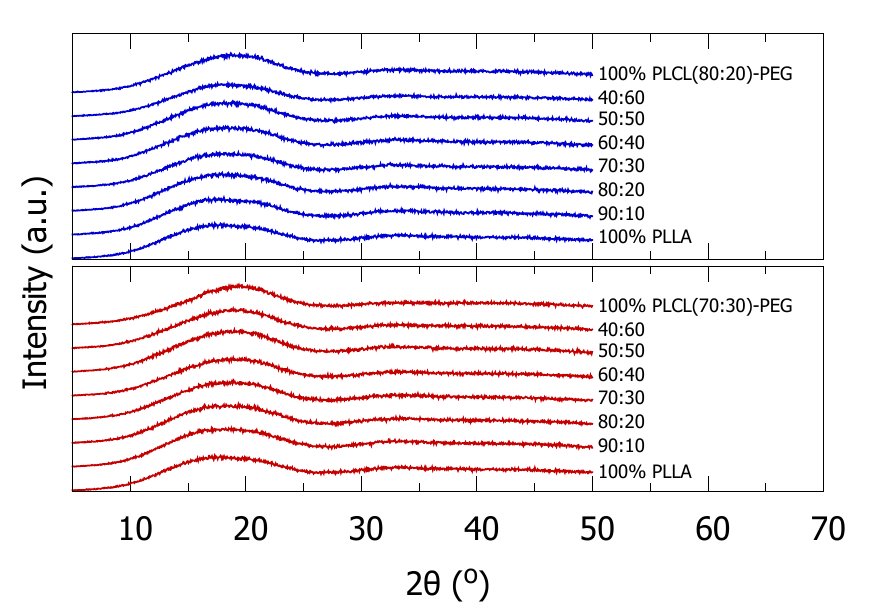}
	\caption{Representative X-ray diffraction patterns for PLLA:PLCL-PEG blends after moulding, labels denote PLLA:PLCL-PEG blend ratio, and datasets are offset for clarity.}
	\label{fig:XRD_t0}
\end{figure}

\vspace{4cm}

\begin{flushleft}
	
	Published journal article:\\
	\url{https://doi.org/10.1016/j.polymdegradstab.2019.109015}\\
	\vspace{0.5cm}
	Copyright \textcopyright\ \href{https://creativecommons.org/licenses/by-nc-nd/2.0/uk/}{CC-BY-NC-ND}\\
	\vspace{0.2cm}
	\includegraphics[width=0.2\linewidth]{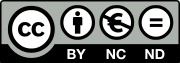}	
	
\end{flushleft}

\end{document}